# Boron-assisted stabilization of low-resistivity mixed-valence Cu–O thin films prepared by reactive magnetron sputtering

Nirmal Kumar[a], Jemal Yimer Damte[a], Michal Procházka[b], Radomír Čerstvý[a], Jiří Houška[a], Pavel Baroch[a], Stanislav Haviar[a], Jiří Rezek[a, *]

[a] Department of Physics and NTIS-European Centre of Excellence, University of West Bohemia in Pilsen, Univerzitní 8, Plzeň, 301 00, Czech Republic.

[b] New Technologies-Research Centre, University of West Bohemia in Pilsen, Univerzitní 8, Plzeň, 301 00, Czech Republic.

* Corresponding author, Tel.: +420 377632269, E-mail address: jrezek@fav.zcu.cz

## Abstract

This study systematically investigated the influence of boron incorporation in Cu-O thin films and the effect of oxygen partial pressure ($p_{ox}$) on the phase evolution, chemical bonding, and electrical characteristics of the prepared films. A phase transition from $Cu_2O$ to $Cu_2O/Cu_4O_3$ to CuO was observed as oxygen partial pressure increased. Boron incorporation significantly broadened the stability window of the $Cu_2O$ and $Cu_4O_3$ phases and delayed the transition to CuO at higher oxygen partial pressure. In the highly B-doped Cu–O films, $Cu_4O_3$ was stabilized even under oxygen-rich conditions along with CuO phase, suggesting that boron significantly altered the oxidation pathway. The formation of B–O and possible B–O–Cu configurations altered the local oxygen chemistry and promoted mixed-valence copper oxide phases. Electrical measurements revealed that highly B-doped Cu–O films exhibited a delayed transition from a high-resistivity low-$p_{ox}$ regime to a low-resistivity mixed-valence regime, ultimately reaching ~0.06 Ω cm, among the lowest reported resistivities for a CuO-like material. These findings demonstrate that boron doping is an effective approach for tailoring the phase stability, defect chemistry, and electrical characteristics of Cu–O thin films for optoelectronic and photovoltaic applications.

1. Introduction

Copper oxide thin films have attracted significant attention owing to their natural abundance, low toxicity, and prospective applications in photovoltaics, transparent electronics, gas sensing, catalysis, and electrochemistry [1–6]. Copper exhibits multiple stable and metastable oxide phases, including cuprous oxide ($Cu_2O$, $Cu^+$), cupric oxide (CuO, $Cu^{2+}$), and mixed-valence paramelaconite ($Cu_4O_3$, $Cu^+/Cu^{2+}$). These distinct phases possess unique electronic configurations and optical characteristics, rendering copper oxides as multifaceted functional materials [7–12]. $Cu_2O$ is a p-type semiconductor with a relatively wide band gap (~2.0–2.6 eV), whereas CuO exhibits a narrower band gap (~1.2–2.1 eV) and stronger defect-mediated electrical conduction [12–14]. Among these phases, $Cu_4O_3$ is particularly interesting because of its mixed-valence nature, containing both $Cu^+$ and $Cu^{2+}$ sites, which may provide localized intervalence charge-transfer pathways and contribute to defect-assisted transport [15]. Nevertheless, beyond phase identification, achieving reliable control over the electrical resistivity remains a key challenge for the practical integration of copper oxide thin films into functional devices. In p-type CuO, the mechanism of electrical conduction is predominantly governed by holes linked to copper vacancies, whereas the carrier mobility is profoundly affected by the microstructural attributes and defect chemistry. Grain boundary potential barriers, ionized impurity scattering, and complexes associated with oxygen-related defects can significantly hinder charge transport in polycrystalline films [16–18]. The oxygen partial pressure ($p_{ox}$) during the reactive magnetron sputtering process critically influences phase formation and the defect concentration [1,16,19–21]. Although higher $p_{ox}$ typically leads to the formation of fully oxidized CuO, excessive oxygen incorporation can introduce interstitial defects and compensation effects that reduce hole mobility, resulting in non-monotonic resistivity trends [1,20]. Therefore, achieving low resistivity in Cu–O thin films requires control of the oxygen chemical potential, defect density, and microstructural characteristics.

Among the various preparation techniques, reactive magnetron sputtering is one of the most versatile for preparing Cu–O thin films because it enables precise control over film composition, phase formation, and microstructure by tuning the oxygen partial pressure during deposition [22,23]. However, controlling the electrical resistivity of Cu–O films remains a major challenge. The resistivity is highly sensitive to oxygen stoichiometry, defect concentration, phase composition, and microstructure. In many cases, highly oxidized CuO-rich films exhibit increased resistivity due to reduced carrier mobility and defect compensation effects, limiting their practical application in oxide electronic devices [24,25]. Doping has emerged as a potent approach for modulating transport properties beyond the oxygen stoichiometry. Light-element dopants, such as boron, are particularly interesting because of their small atomic radius and strong affinity for oxygen, which can influence the formation energies of Cu vacancies, alter the heights of grain boundary barriers, and modify local redox environments [22,26,27]. Numerous studies have been extensively explored the influence of various dopants on the structural, optical, and electrical properties of copper oxide systems [22,26,28–30]. Nevertheless, the effects of boron doping on Cu–O thin films deposited under systematically varied oxygen partial pressures have not been comprehensively examined. In particular, the role of boron in altering phase stability and facilitating defect-assisted transport under oxygen-rich growth conditions remains largely unexplored.

In this work, undoped and B-doped Cu–O thin films were deposited by reactive magnetron sputtering under varying oxygen partial pressures. The influence of boron incorporation and $p_{ox}$ on phase evolution, microstructure, chemical bonding, optical band gap, and electrical charge transport was systematically investigated using X-ray diffraction (XRD), Raman spectroscopy, scanning electron microscopy (SEM), energy-dispersive X-ray spectroscopy (EDS), X-ray photoelectron spectroscopy (XPS), Hall measurements, and density functional theory (DFT) calculations. The main focus is on understanding the origin of the remarkably low resistivity in highly B-containing Cu–O films, where CuO coexists with mixed-valence and boron–oxygen-related local environments.

## 2. Experimental setup

*Materials synthesis*

The thin films were deposited on silicon and soda-lime glass substrates. Prior to deposition, the substrates were cleaned sequentially in an ultrasonic bath with deionized water and isopropyl alcohol for 15 min each. After cleaning, the substrates were dried using high-purity nitrogen gas and immediately loaded into the deposition chamber. Thin films were deposited using a reactive magnetron sputtering system. Three circular sputtering targets with a diameter of 100 mm and a thickness of 6 mm (area of 78.54 $cm^2$) were employed: pure Cu, 0.5 wt.% B-doped Cu, and 1.5 wt.% B-doped Cu, all with the purity of 99.99%. The distance between the target and substrate was 100 mm. The sputtering chamber was evacuated using a diffusion pump backed by a rotary pump, achieving a base pressure of approximately $3\times10^{-3}$ Pa. Argon was used as the sputtering gas, and its working pressure was maintained at 0.5 Pa with constant pumping speed throughout all depositions. Oxygen was introduced as a reactive gas, and the oxygen partial pressure was varied from 0.10 to 1.0 Pa to control the film composition. The substrate holder temperature was 200 °C for all experiments. The average target power was fixed at 500 W, corresponding to a pulse-averaged target power density of 12.75 $Wcm^{-2}$, by setting $t_{on} = t_{off} = 20$ µs. The deposition time was between 120 s and 360 s, resulting in a film thickness of 130-250 nm. All other deposition parameters were kept constant for all samples to ensure reproducibility. For clarity, the films prepared using "pure Cu target" are denoted as "undoped films", the films prepared using "0.5 wt.% boron in Cu target" are denoted as "lightly B-doped Cu–O" films, and those prepared using "1.5 wt.% boron in Cu target" are denoted as "highly B-doped Cu–O" films.

*Characterization techniques*

The structural and phase properties of the deposited films were investigated by X-ray diffraction (XRD) using PANalytical X'Pert PRO diffractometer working in the Bragg-Brentano geometry using a CuKα (40kV, 40mA) radiation and Raman spectroscopy (Horiba Jobin Yvon LABRAM HR Evolution Raman Spectroscope). The surface morphology and grain structure were examined using scanning electron microscopy (SEM, SU-70, Horiba Ltd., Japan) with a primary electron energy of 10 keV in top-view mode. At the same time, cross-sectional SEM imaging was employed to confirm film thickness and uniformity.

The X-ray photoelectron spectroscopy (XPS) measurements were done in an ultra-high vacuum (UHV) chamber with base pressure $\leq 4\times10^{-9}$ mbar using a hemispherical analyzer Phoibos 150 (SPECS Surface Nano Analysis GmbH) with a multichannel CMOS detector. Non-monochromatic X-ray source XR 50 was operated with Mg Kα line (1253.6 eV). The survey spectra were obtained by 5 scans with an energetic step size of 0.5 eV and a pass energy of 50 eV. Core-level spectra were measured with 20 scans, a step size of 0.1 or 0.05 eV, and a pass energy of 30 eV. Samples were mounted on Ta or Mo sample holders using carbon tape and silver paste for conducting connections. All spectra were charge corrected based on the adventitious C 1s spectral component (C–C, C–H) with a binding energy (BE) of 284.8eV. KolXPD software was used to analyze the measured spectra with Shirley or linear background, and with Voigt and Voigt doublet functions.

The elemental composition of the films was analyzed using energy-dispersive X-ray spectroscopy (EDS, Oxford Instruments, Max Infinity) at a primary beam energy of 4 keV. Quantification was performed using $Cu_2O$ (Astimex) and B (MAC) standards. The influence of the substrate on the measured composition was evaluated using the thin-film analysis module (LayerProbe®). The trueness of the analysis was estimated to be approximately 1, 2, and 0.5 at.% for Cu, O, and B, respectively.

The Hall effect measurements (MMR Technologies) were performed to determine the carrier mobility, carrier concentration, electrical resistivity, and conduction type of the films. The electrical resistivities of the films were also confirmed using a standard four-point probe (4PP) method. The film thickness was measured using a stylus profilometer (Dektek 8 Profiler) and cross-verified by SEM cross-sectional images obtained with secondary and backscattered electrons. Optical band gap, $E_g$, was determined by using Tauc plot method. The film transmittance, *T*, and reflectance, *R*, at the angle of 7◦ from the sample normal were measured at room temperature using the Agilent CARY 7000 spectrophotometer. The absorption coefficient α is calculated as $\alpha = -\left[ln\frac{T}{1-R}\right]\frac{1}{t}$ , where *t* is the thickness of the thin film layer. The direct band gap for CuO films was calculated using the relationship $(\alpha h\nu)^2 = (h\nu - E_g)$, where $h\nu$ is the photon energy. A plot of $(\alpha h\nu)^2$ versus $h\nu$ was used to determine the $E_g$.

*Computational details*

Density functional theory (DFT) calculations were performed using the Vienna *ab initio* Simulation Package (VASP) [31]. The exchange–correlation effects were described within the generalized gradient approximation using the Perdew–Burke–Ernzerhof functional [32], while the interaction between core and valence electrons was treated using the projector augmented wave approach [33]. To improve the description of the localized Cu 3d electrons and better reproduce the electronic structure of CuO and $Cu_2O$, we applied the Hubbard U correction (DFT+U) with an effective $U_{\mathrm{eff}} = 4.5$ eV on Cu 3d states. A similar value has been used in previous first-principles studies of CuO and $Cu_2O$, where a self-consistent of U 4.5 eV was found to provide a reasonable description of the electronic structure [34]. A 2 × 2 × 2 supercell of CuO (64 atoms) and $Cu_2O$ (48 atoms) was constructed to model the bulk structure and defect configurations. The plane-wave basis set was expanded to a kinetic energy cutoff of 550 eV, and Brillouin-zone integrations were performed using a Monkhorst–Pack k-point mesh of $6 \times 6 \times 6$.

All atomic positions and lattice degrees of freedom were fully optimized until the residual forces on each atom were below 0.005 eV $Å^{-1}$, and the total energy convergence criterion was set to $1 \times 10^{-6}$ eV to ensure high numerical accuracy. Long-range dispersion interactions, which are not adequately captured by standard GGA functionals, were included using the DFT-D3 correction scheme [35]. Boron doping in $Cu_2O$ was investigated by considering both interstitial and substitutional defect configurations. In particular, boron atoms were introduced into the interstitial sites and substituted at the Cu or O lattice positions **Figure 1a and b**. The energies of Cu or O vacancy formation, both with and without the presence of B, were examined. The relative stability of these defect structures was assessed by calculating their formation energies using the total energies obtained from the fully relaxed geometries. An energy correction of 1.36 eV was applied to the $O_2$ molecule to compensate for the GGA overbinding and correlation errors (following the approach proposed by Wang et al. [36]). Energies of all oxides were adjusted by -1.90 eV per Cu atom in order to reproduce the experimental [37] formation energies of pure stoichiometric CuO and $Cu_2O$ as demonstrated in **Figure 1c** (following the approach proposed for mixing DFT and DFT+U by Jain et al. [38], albeit the value of the correction term has been recalculated in order to correspond to the present methodology, in the first place the U value).

For boron interstitials, the formation energy was calculated as

$$E_{\mathrm{f}}^{\mathrm{int}} = E_{\mathrm{tot}}\,(Cu_2O + B) - E_{\mathrm{tot}}\,(Cu_2O) - \mu_{\mathrm{B}}$$

Where $E_{\mathrm{tot}}$ ($Cu_2O$ + B) and $E_{\mathrm{tot}}$ ($Cu_2O$) are the total energies of the supercell with and without the boron atom, respectively, and $\mu_{\mathrm{B}}$ is the chemical potential of boron, taken from bulk boron.

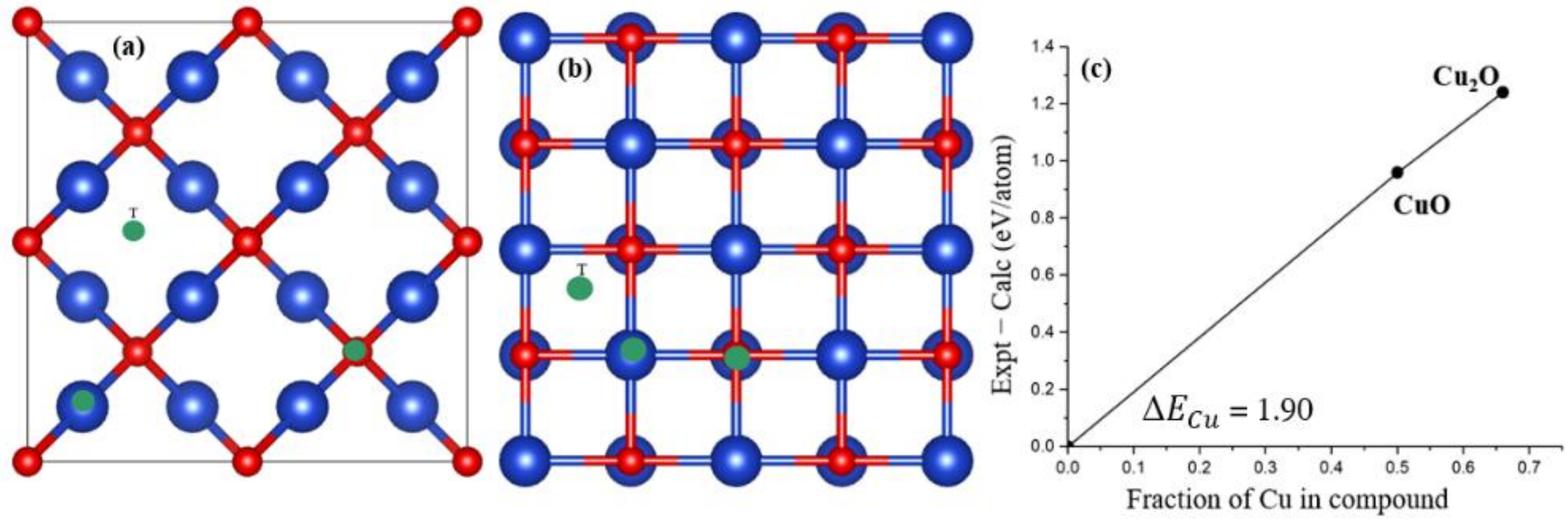

Figure 1 Optimized structures of (a) $Cu_2O$, (b) CuO and (c) difference between experiment [37] and calculated formation energies (Expt − Calc, eV/atom) as a function of the copper fraction in CuO compounds. The linear trend obtained from CuO and $Cu_2O$ yields a correction parameter of $\Delta E_{\mathrm{Cu}}$ = 1.90 eV, which is used to account for systematic DFT errors in copper-containing oxides. Copper (Cu) atoms are shown in blue, oxygen (O) atoms in red, and boron (B) atoms in green.

For substitutional boron at a Cu site, the formation energy was computed using

$$E_{\mathrm{f}}^{\mathrm{sub}} = E_{\mathrm{tot}}\,(Cu_2O{:}B_{\mathrm{Cu}}) - E_{\mathrm{tot}}\,(Cu_2O) - \mu_{\mathrm{B}} + \mu_{\mathrm{Cu}}$$

where $E_{\mathrm{tot}}$ ($Cu_2O$ : $B_{\mathrm{Cu}}$) represents the total energy of the supercell with boron replacing a Cu atom, and $\mu_{\mathrm{Cu}}$ is the chemical potential of copper, obtained from bulk Cu.

Additionally, the formation energy of a Cu vacancy was determined as

$$E_{\mathrm{f}}^{V_{\mathrm{Cu}}} = E_{\mathrm{tot}}\,(Cu_2O{:}V_{\mathrm{Cu}}) - E_{\mathrm{tot}}\,(Cu_2O) + \mu_{\mathrm{Cu}}$$

Where = $E_{\mathrm{tot}}$ ($Cu_2O$:$V_{\mathrm{Cu}}$) is the total energy of the supercell containing a Cu vacancy. Defect formation energies in CuO were calculated using the same expressions as those employed for $Cu_2O$, replacing the corresponding $Cu_2O$ total energies with those of CuO.

3. Results and Discussion

3.1. DFT Calculations

**Table 1** summarizes the calculated formation energies of boron-related point defects and defect vacancy complexes in CuO and $Cu_2O$, clearly indicating that defect stability is highly sensitive to the local defect environment.

Table 1 Calculated formation energies $E_f$ (eV) of boron-related point defects and defect vacancy complexes in CuO and $Cu_2O$. $B_O$, $B_{Cu}$ and $B_i$ represent boron substituting oxygen, boron substituting copper, and boron at an interstitial site, respectively. Vacancy-assisted configurations correspond to copper ($V_{Cu}$) and oxygen ($V_O$) vacancies located either near or far from the boron dopant.

| Defect type (B) | CuO | CuO + V (far B) | CuO + V (near B) | $Cu_2O$ | $Cu_2O$ + V (far B) | $Cu_2O$ + V (near B) |
|---|---|---|---|---|---|---|
| $B_O$ | 2.31 | 2.42 ($V_{Cu}$) | 4.72 ($V_{Cu}$) | 6.10 | 7.84 ($V_{Cu}$) | 7.88 ($V_{Cu}$) |
| | | 4.28 ($V_O$) | 4.33 ($V_O$) | | 9.25 ($V_O$) | 9.07 ($V_O$) |
| $B_{Cu}$ | -0.15 | -2.19 ($V_{Cu}$) | -2.83 ($V_{Cu}$) | 3.10 | 3.46 ($V_{Cu}$) | 1.99 ($V_{Cu}$) |
| | | -1.88 ($V_O$) | -1.34 ($V_O$) | | 6.54 ($V_O$) | 7.83 ($V_O$) |
| $B_i$ | -4.79 | -2.99 ($V_{Cu}$) | - | 3.76 | 4.93 ($V_{Cu}$) | - |
| | | -2.95 ($V_O$) | | | 6.34 ($V_O$) | |
| Vacancy | 0.68 ($V_{Cu}$) | - | - | 2.27 ($V_{Cu}$) | - | - |
| | 0.94 ($V_O$) | | | 3.25 ($V_O$) | | |

In CuO, interstitial boron ($B_i$) exhibits the lowest formation energy (−4.79 eV) under stoichiometric conditions, indicating that interstitial incorporation is intrinsically favorable. Substitutional boron at the Cu site ($B_{Cu}$) is slightly less stable (−0.15 eV), while substitution at the oxygen site ($B_O$) is energetically unfavorable (2.31 eV). The presence of copper or oxygen vacancies significantly modifies the stability. In particular, $B_{Cu}$ coupled with a nearby copper vacancy yields strongly negative formation energies (−2.83 eV), demonstrating that Cu-deficient conditions promote boron incorporation at Cu sites. In contrast, vacancy-assisted $B_O$ configurations remain energetically costly. For $Cu_2O$, the trends differ markedly. Under stoichiometric conditions, substitutional boron at the Cu site (3.10 eV) and interstitial boron (3.76 eV) are both energetically unfavorable, while $B_{\mathrm{O}}$ exhibits even higher formation energies (6.10 eV). However, the stability improves in the presence of copper vacancies, particularly for $B_{\mathrm{Cu}}$ associated with a nearby $V_{\mathrm{Cu}}$ (1.99 eV), indicating that vacancy-assisted incorporation is energetically more accessible in $Cu_2O$ under Cu-poor conditions. In general, CuO shows a greater intrinsic tendency to incorporate boron than $Cu_2O$, especially at interstitial and Cu-substitutional sites. The significant stabilization observed in vacancy-assisted configurations in both lattices highlights the critical role of copper-deficient environments in facilitating boron doping. These energetically favorable configurations were therefore selected for further electronic structure analysis.The electronic structures of pristine and defect-containing CuO and $Cu_2O$ were examined using the calculated total density of states (DOS) shown in **Figure 2**.

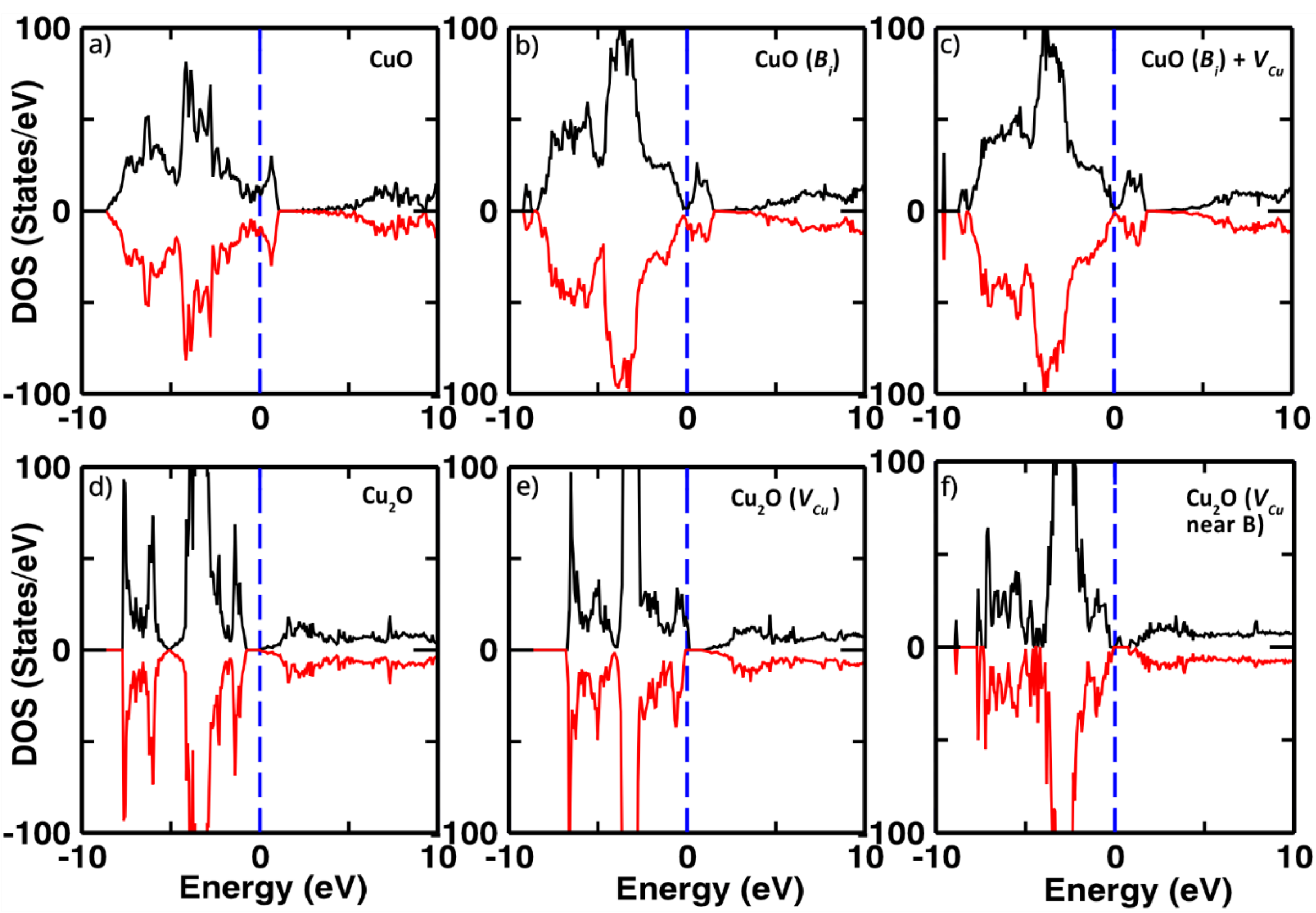

Figure 2 Total density of states (DOS) of (a) pristine CuO, (b) CuO with boron at an interstitial site, (c) CuO with a copper ($V_{\mathrm{Cu}}$) vacancies located far from the boron dopant at an interstitial site, (d) pristine $Cu_2O$, (e) $Cu_2O$ with a copper ($V_{\mathrm{Cu}}$) vacancy and (f) $Cu_2O$ with a copper ($V_{\mathrm{Cu}}$) vacancies located near the boron dopant. The Fermi level is aligned at 0 eV.

Pristine CuO exhibits a well-defined band gap of approximately 1.6 eV. The introduction of boron at an interstitial site significantly narrows the band gap to about 0.90 eV. When a copper vacancy is positioned far from the boron dopant, the band gap remains reduced (≈1.03 eV), indicating that defect interactions strongly influence the electronic structure. This reduction arises from the appearance of defect-induced states near the band edges, reflecting enhanced electronic coupling between boron and copper vacancies in the CuO lattice. A similar behavior is observed in $Cu_2O$, although the overall band-gap modification is less pronounced. The pristine $Cu_2O$ system shows a band gap of approximately 1.2 eV, which decreases to about 1.06 eV when a copper vacancy is located near the boron dopant. In the presence of a copper vacancy, located far from boron, the band gap further reduces slightly to around 1.04 eV. These results demonstrate that vacancy-assisted boron incorporation modifies the electronic structure of $Cu_2O$ by introducing additional states near to the Fermi level. Boron-doped CuO exhibits a narrower band gap and a higher density of states near the Fermi level compared with the corresponding $Cu_2O$ configurations, as clearly shown in **Figure 2.** The increased electronic states near the Fermi level suggest enhanced charge carrier availability in CuO. Consequently, boron-doped CuO is expected to exhibit higher electrical conductivity than $Cu_2O$, although a quantitative evaluation would require explicit transport calculations.

Table 2 Bader charges (in |e|) of nearest-neighbor Cu and O atoms for boron-doped CuO and $Cu_2O$ with a nearby copper vacancy.

| Species | Cu1 | Cu2 | O1 | O2 |
|---|---|---|---|---|
| $Cu_2O$ + $V_{\mathrm{Cu}}$ (near B) | -0.20 | -0.24 | -0.69 | -0.94 |
| CuO + $V_{\mathrm{Cu}}$(near B) | 0.27 | 0.29 | -0.59 | -0.57 |

**Table 2** shows, that for the copper-vacancy–boron complex with the vacancy located near the dopant, CuO exhibits significantly larger charge depletion on neighboring Cu atoms compared with $Cu_2O$, accompanied by notable charge accumulation on surrounding O atoms. This enhanced charge redistribution indicates stronger electronic delocalization in CuO, which is consistent with the higher density of states near the Fermi level observed in **Figure 2**. In contrast, the more localized charge distribution in $Cu_2O$ correlates with its lower DOS near the Fermi level, confirming that boron–vacancy interactions are electronically more active in CuO. As a result, boron-doped CuO is expected to exhibit higher electrical conductivity than the corresponding $Cu_2O$ system.

3.2. Deposition rate and film structure

The deposition rate decreases significantly with increasing $p_{ox}$ for all the films, as shown in **Figure 3**, which is characteristic of the reactive sputtering process due to increasing target poisoning and reduced sputtering yield under an oxygen-rich environment [39]. At low $p_{ox}$, the films exhibit relatively high deposition rates because the target surface

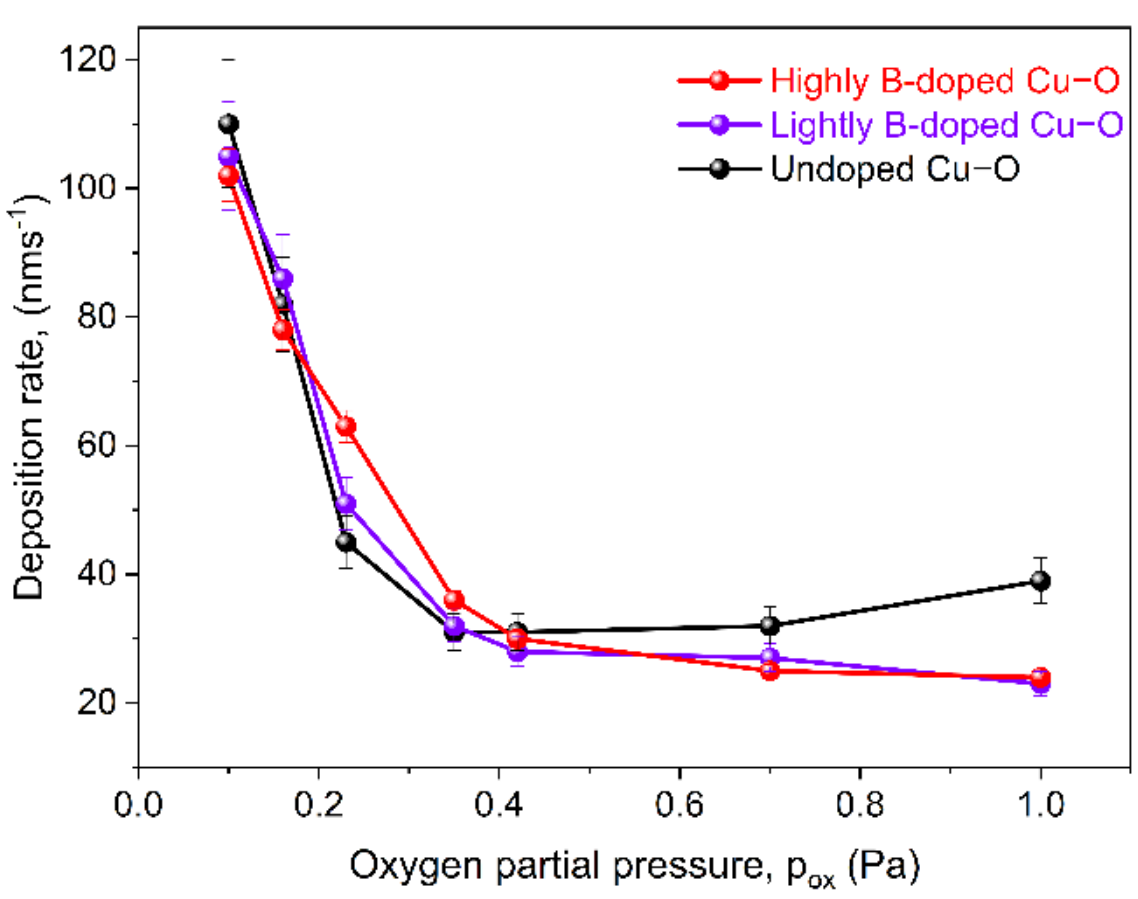

Figure 3 Variation of the deposition rate of undoped and B-doped Cu-O thin films as a function of oxygen partial.

remains mostly metallic. As the $p_{ox}$ increases, oxidation of the target surface reduces sputtering

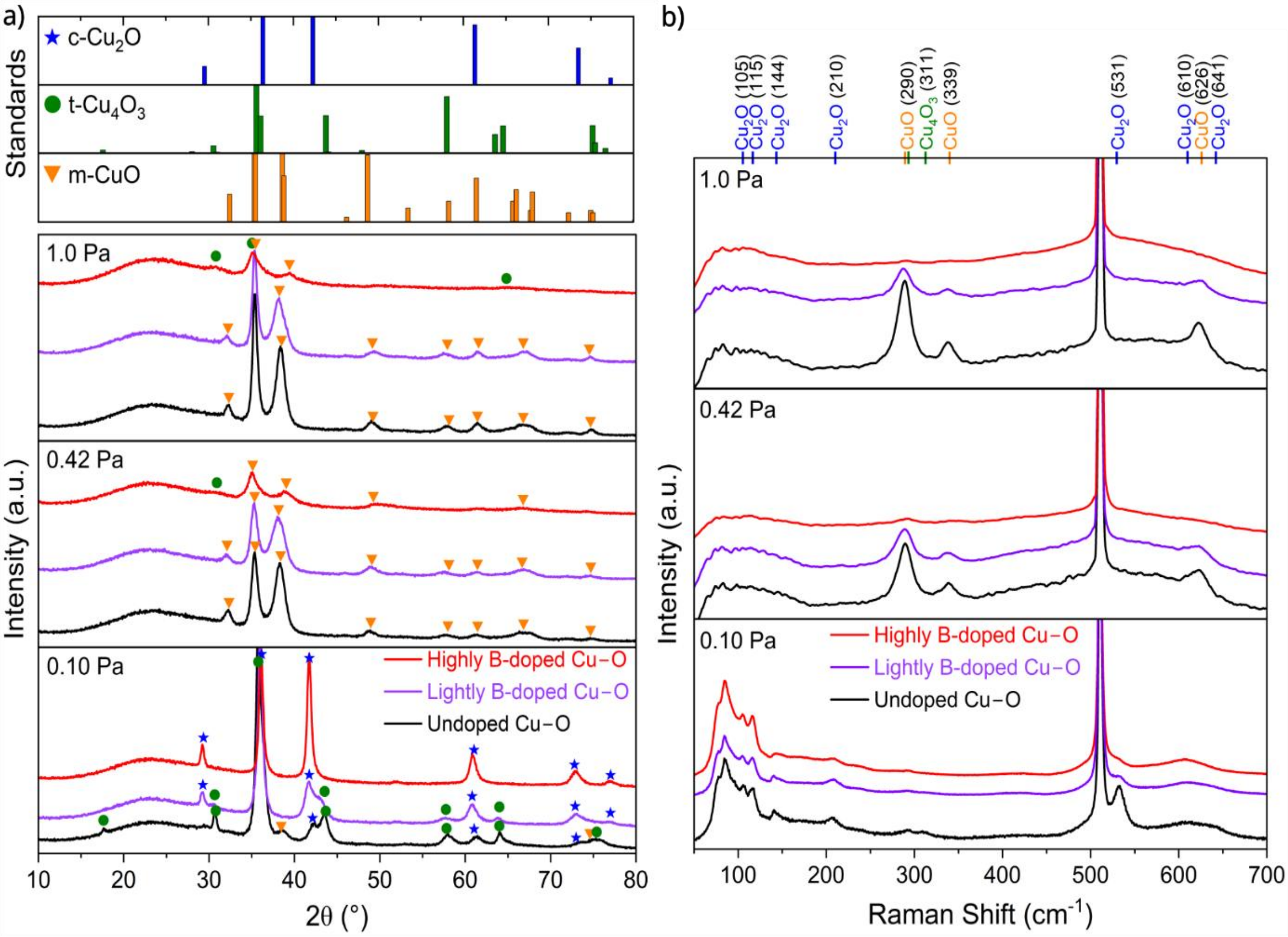

Figure 4 XRD patterns (a) and Raman spectra (b) of undoped and B-doped Cu–O thin films deposited at selected oxygen partial pressures ($p_{ox}$).

efficiency, leading to a sharp decrease in the deposition rate.

The crystalline structure and phase evolution of the deposited films depend on variations in oxygen partial pressure and boron concentration and are examined using XRD and Raman spectroscopy, with the resulting results illustrated in **Figure 4a**, and **Figure 4b**, respectively. The standard diffraction patterns for cubic $Cu_2O$ (PDF Card No. 00-005-0667), tetragonal $Cu_4O_3$ (PDF Card No. 04-007-2184), and monoclinic CuO (PDF Card No. 00-048-1548) are shown in **Figure 4a** for comparative analysis. The XRD patterns clearly demonstrate that both the oxygen partial pressure and the boron concentration significantly influence the phase formation of the films.

At a low oxygen partial pressure of 0.10 Pa, notable discrepancies in phase composition are observed as a function of boron concentration. For the undoped film, the diffraction pattern revealed the coexistence of $Cu_4O_3$ and $Cu_2O$, with $Cu_4O_3$ as the predominant phase and $Cu_2O$ as a secondary phase. In lightly B-doped Cu–O films, the relative phase composition is altered such that $Cu_2O$ is a dominant phase, with $Cu_4O_3$ as a secondary phase, thereby indicating that the presence of boron modifies the oxidation pathway throughout the film growth process. A further increase in boron concentration (highly B-doped Cu–O) results in the predominant formation of the $Cu_2O$ phase at the identical oxygen partial pressure, characterized by the total dominance of $Cu_2O$ reflections within the XRD pattern, while the $Cu_4O_3$ phase is substantially suppressed. This behavior suggests that an elevated boron concentration preferentially stabilizes the $Cu^+$ oxidation state under conditions of oxygen deficiency during growth [9,22,40].

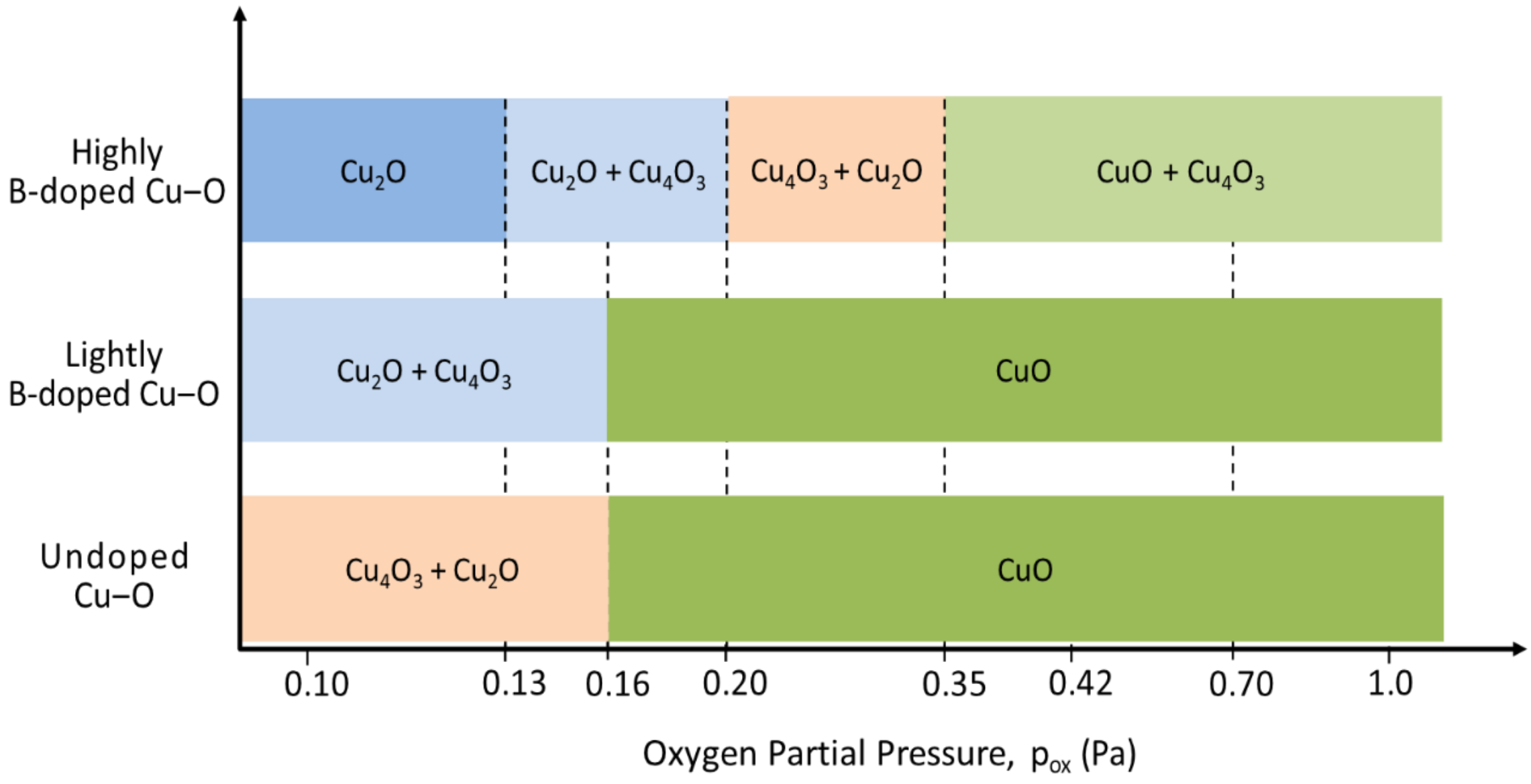

Figure 5 Schematic illustration of the phase evolution of undoped and B-doped Cu–O thin films as a function of oxygen partial pressure. The diagram shows the transition between $Cu_2O$, $Cu_4O_3$, and CuO phases and highlights the broadening of the mixed-phase region and delayed CuO formation with increasing boron concentration.

As the oxygen partial pressure increases, the films transition to more oxidized phases, although the transition pressure depends strongly on the boron concentration [41]. For the undoped Cu–O films, the transition from sub-stoichiometric oxides to fully oxidized CuO occurs at a relatively low oxygen partial pressure. In this instance, the XRD patterns indicate that the films' transition to the CuO phase begins at approximately 0.16 Pa and remains the dominant phase up to 1.0 Pa. Conversely, when boron is incorporated into the films, the transition from the $Cu_2O$ or $Cu_2O/Cu_4O_3$ phases to CuO occurs at 0.16 Pa and 0.35 Pa for lightly and highly B-doped films, respectively, thereby indicating that boron widens the stability window of the lower-valence copper oxide phases [22,41].

For the highly B-doped Cu–O films, the phase evolution with oxygen partial pressure proceeds through several intermediate stages (**Figure 5**). At $p_{ox}$ = 0.16 Pa, the diffraction pattern indicates a mixed-phase where $Cu_2O$ remains the primary phase, and $Cu_4O_3$ appears as a secondary phase. Increasing the oxygen partial pressure to 0.20 Pa further shifted the phase balance, with $Cu_4O_3$ becoming the dominant phase while $Cu_2O$ began to disappear. Upon further increasing the oxygen partial pressure to 0.35 Pa, the diffraction patterns indicate the presence of multiple phases: monoclinic CuO + $Cu_4O_3$ in all films, and this remains consistent up to 1.0 Pa, as shown in **Figure 5**. At higher oxygen partial pressures, the phase composition becomes dependent on the boron concentration: at $p_{ox}$ = 0.70 Pa, the undoped and lightly B-doped films remain dominated by CuO reflections, indicating that CuO remains the stable phase under these oxygen-rich conditions. However, for the highly B-doped Cu–O films, additional diffraction peaks corresponding to $Cu_4O_3$ appear alongside the CuO reflections, indicating the formation of a mixed $CuO/Cu_4O_3$ phase. Even at the highest oxygen partial pressure of 1.0 Pa, this effect remains uniform in the highly B-doped films. In addition to the phase evolution, slight shifts in the diffraction peak positions are observed with increasing boron concentration, indicating modifications in the local lattice environment of the Cu–O films. Such peak shifts are commonly associated with lattice distortion, defect formation, and local strain induced by dopant incorporation and non-stoichiometry.

The phase evolution observed in the XRD measurements was further supported by the Raman spectroscopy results, as shown in **Figure 4b**. The Raman spectra collected at representative oxygen partial pressures exhibited vibrational modes characteristic of the corresponding copper oxide phases. At $p_{ox}$ = 0.10 Pa, the spectra show contributions from both $Cu_2O$ the and $Cu_4O_3$ phases. For the undoped film, the Raman features associated with $Cu_4O_3$ were more pronounced, consistent with the XRD results, which showed $Cu_4O_3$ as the dominant phase. With increasing boron concentration, the relative intensity of the Raman modes associated with $Cu_2O$ increased, confirming the stabilization of this phase. For the highly B-doped Cu–O film, the Raman spectrum is dominated by $Cu_2O$-related modes, supporting the XRD observation of a predominantly $Cu_2O$ structure at this oxygen partial pressure.

At $p_{ox}$ = 0.42 Pa, the Raman spectra for all films are dominated by the characteristic vibrational modes of CuO, with a strong peak observed near ~530 $cm^{-1}$ corresponding to the $A_g$ vibrational mode of monoclinic CuO. At higher oxygen partial pressures, particularly at $p_{ox}$ = 1.0 Pa, the Raman spectra of the undoped and lightly B-doped Cu–O films are dominated by CuO modes. However, the Raman spectrum signal from the highly B-doped films is not intense enough to identify the peaks corresponding to the $Cu_4O_3$ phase.

The combined XRD and Raman analyses revealed a systematic evolution of copper oxide phases as a function of the oxygen partial pressure and boron doping level (**Figure 5**). In the case of undoped films, the transition from $Cu_2O/Cu_4O_3$ to CuO occurred at a relatively low oxygen partial pressure (~0.13 Pa). However, with increasing boron concentration, the transition to CuO shifted to higher oxygen partial pressures, indicating a broadening of the stability window of the $Cu_2O$ and $Cu_4O_3$ phases. This behavior demonstrates that boron incorporation significantly alters the oxidation dynamics and phase stability of copper oxide thin films, enabling tunability of the structural phases over a broader range of oxygen partial pressures. First-principles calculations reveal that boron incorporation becomes significantly more stable in the presence of Cu vacancies, especially in CuO-based structures.

### 3.3. XPS and EDS analysis

The chemical states of Cu, O, and B in the Cu–O-B thin films were analyzed using XPS. Since XPS is very surface sensitive, all samples were measured in the as-introduced state (not shown here) as well as after 3 minutes of sputtering in an Ar atmosphere in order to obtain the signal from the real samples, not from surface oxides and contaminants created on top due to the transfer under ambient pressure.

The Cu 2p spectra, illustrated in **Figure 6**, show characteristic Cu $2p_{3/2}$ and Cu $2p_{1/2}$ peaks, with deconvolution revealing contributions from both $Cu^{+}$ (~932.2–932.8 eV) and $Cu^{2+}$ (~933.5–

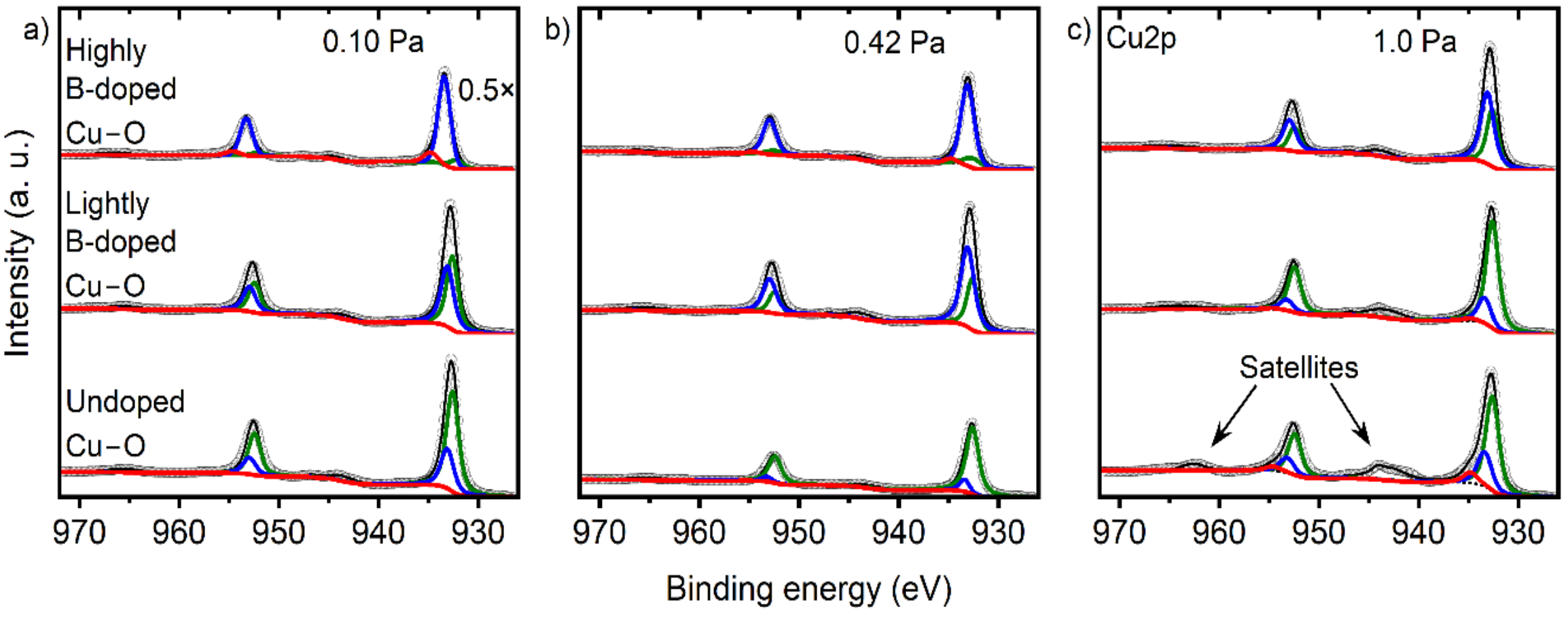

Figure 6 High-resolution Cu 2p XPS spectra of undoped and B-doped Cu–O thin films deposited at oxygen partial pressure, (a) 0.10 Pa, (b) 0.42 Pa, and (c) 1.0 Pa. The deconvoluted spectra indicate the coexistence of $Cu^{+}$ (blue curve) and $Cu^{2+}$ (green curve) states and show the evolution from $Cu_2O$/$Cu_4O_3$-rich to CuO-rich phases with increasing oxygen partial pressure.

934.8 eV) states corresponding to $Cu_2O$ (blue component) and CuO (green component), while red component (~934.8–935.1 eV), noticeable in some spectra correspond to $Cu(OH)_2$ [42,43]. All undoped Cu–O samples exhibit a dominant peak at a lower binding energy, with a small contribution at a higher binding energy, as evidenced by low-intensity satellite peaks (~942–946 eV).

An increase in boron concentration leads to a shift to higher binding energy, and $Cu^{2+}$ becomes dominant at $p_{ox}$ = 0.10 Pa and 0.42 Pa, respectively. For $p_{ox}$ = 1.0 Pa, the shift to higher binding energy is less pronounced, and the highly-B-doped sample still exhibits clear coexistence of $Cu^{+}$ and $Cu^{2+}$ states, suggesting stabilization of a mixed-valence environment. However, despite the apparent increase in the CuO-related contribution, the satellite peaks remain weak or significantly suppressed compared with those typically reported for stoichiometric CuO, which is generally characterized by intense satellite features [43,44]. Li et al. [29] reported that boron facilitates the transfer of Cu to higher binding-energy levels. Furthermore, Wang et al. [14] claim that the $Cu_2O$ and $Cu_4O_3$ phases may overlap with the CuO peak, and that the satellite peak of $Cu_4O_3$ exhibits a similar but less intense profile compared to CuO. In addition, $Ar^{+}$ bombardment also plays a significant role, disrupting the structure and reducing surface oxides

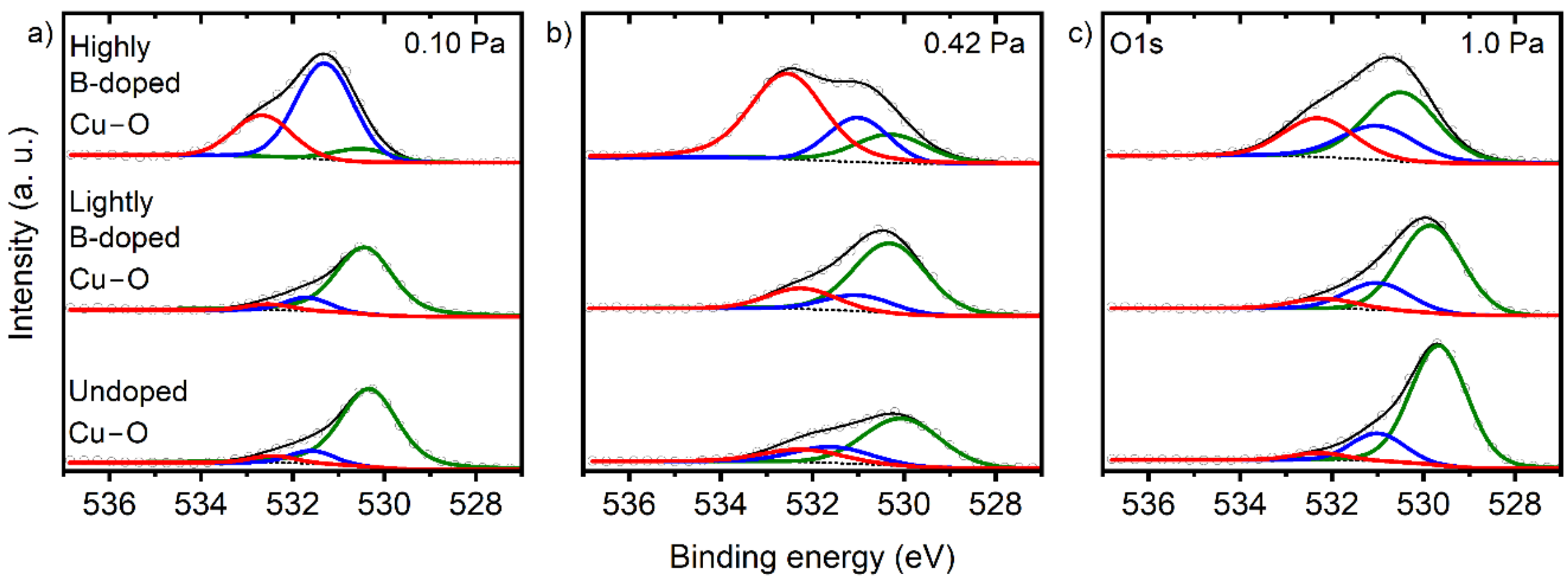

Figure 7 High-resolution O 1s XPS spectra of undoped and B-doped Cu–O thin films deposited at oxygen partial pressure, (a) 0.10 Pa, (b) 0.42 Pa, and (c) 1.0 Pa. The fitted components correspond to lattice oxygen (green component), defect-related oxygen species (blue component), and B–O-related bonding environments (red component).

[44]. Overall, the XPS results suggest that the film surfaces consist of a complex mixture of $Cu_2O$-, $Cu_4O_3$-, and CuO-related phases with varying $Cu^{+}/Cu^{2+}$ ratios.

The O 1s spectra are asymmetric and consist of multiple components corresponding to lattice oxygen (~529.5–530.2 eV), defect-related oxygen (~530.8–531.5 eV), and higher binding energy species (~531.8–533.0 eV) associated with B–O bonding and adsorbed oxygen [15,29,43], which can still be present even after relatively short $Ar^{+}$ bombardment (**Figure 7**).

With increasing boron concentration, the relative intensity of the higher-binding-energy components increases, indicating enhanced oxygen disorder and the formation of boron–oxygen bonds.

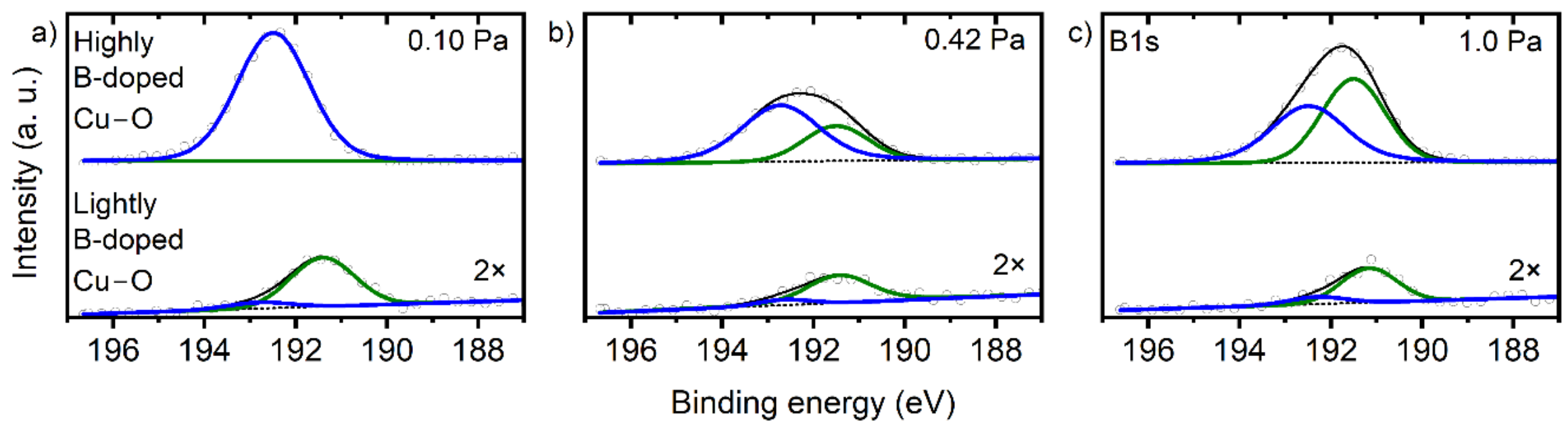

Figure 8 High-resolution B 1s XPS spectra of low and highly B-doped Cu–O thin films deposited at oxygen partial pressure, (a) 0.10 Pa, (b) 0.42 Pa, and (c) 1.0 Pa. The spectra indicate oxidized boron-containing environments, consistent with B–O (blue) bonding and possibly B–O–Cu-like (green) local coordination.

The core-level XPS results indicate that boron is predominantly present in an oxidized B–O bonding state. This is verified by the distinctive B 1s core-level peak centered within the binding energy range, shown in **Figure 8**, typically characteristic of highly oxidized boron environments (~191.0 - 193.5 eV) [42,45,46].

However, while XPS provides localized chemical coordination data, surface photoemission alone cannot unambiguously determine the precise crystallographic position or structural registry of boron atoms within the thin-film matrix. The observed B–O bonding signature can theoretically be attributed to a variety of structural configurations. One possible interpretation is that the B–O signal originates from boron incorporated into the Cu–O matrix, forming local B–O–Cu bonding environments. Alternatively, the signal could stem from dopant segregation at structural grain boundaries or from localized, short-range disordered amorphous $B_2O_3$-like clusters [45,47]. According to DFT calculations, the tendency of boron to create a locally stabilized boron oxide environment is described in the supplementary document (**Figure S1**).

Consequently, the designation “$B_2O_3$-like" must be interpreted as a description of the localized chemical bonding coordination rather than direct proof of a distinct crystalline phase. Because complementary bulk structural diagnostics from X-ray diffraction (XRD) and Raman spectroscopy show no long-range ordered $B_2O_3$ phase reflections, the precipitation of an independent boron oxide secondary phase cannot be conclusively confirmed from the present dataset.

The XPS results indicate that boron is mainly incorporated in oxidized B–O bonding environments and is associated with an increased amount of defect-related oxygen. This modified local oxygen chemistry is consistent with the XRD and Raman results, which show that high boron incorporation shifts the Cu–O phase evolution and promotes the persistence of mixed-valence $Cu_4O_3$ at high $p_{ox}$. DFT calculations support this interpretation by showing that boron-related defects are more favorable in Cu-deficient environments and can introduce electronic states near the Fermi level. Therefore, boron appears to influence both the oxidation pathway and the defect-mediated electronic structure of Cu–O thin films.

Table 3 The film composition at various pox and boron concentrations in the Cu–O films was calculated from EDS spectra.

| $p_{ox}$ (Pa) | Boron concentration | Atomic percentage (at. %) | | | Cu/O ratio |
|---|---|---|---|---|---|
| | | Cu | O | B | |
| 0.10 | Undoped Cu–O | 57.7 | 42.1 | 0 | 1.4 |
| | Lightly B-doped Cu–O | 60.5 | 38.6 | 0.9 | 1.6 |
| | Highly B-doped Cu–O | 57.7 | 38.5 | 3.8 | 1.5 |
| 0.42 | Undoped Cu–O | 48.6 | 51.3 | 0 | 0.95 |
| | Lightly B-doped Cu–O | 48.2 | 50.9 | 0.9 | 0.95 |
| | Highly B-doped Cu–O | 44.0 | 52.4 | 3.6 | 0.84 |
| 1.0 | Undoped Cu–O | 49.3 | 50.7 | 0 | 0.97 |
| | Lightly B-doped Cu–O | 48.3 | 51.0 | 0.7 | 0.95 |
| | Highly B-doped Cu–O | 43.0 | 51.5 | 5.5 | 0.83 |

EDS analysis, shown in **Table 3** confirmed successful incorporation of boron into the Cu–O films, with the boron concentration increasing from approximately 0.7–0.9 at.% in the lightly B-doped films to 3.6–5.5 at.% in the highly B-doped films. At low oxygen partial pressure ($p_{ox}$=0.10 Pa), all films exhibited Cu-rich compositions (Cu/O=1.4–1.6), consistent with the formation of $Cu_2O$- and $Cu_4O_3$-rich phases observed by XRD. Increasing the $p_{ox}$ to 0.42 and 1.0 Pa resulted in a significant increase in oxygen incorporation, reducing the Cu/O ratio to approximately 0.95 for the undoped and lightly B-doped films and to 0.83–0.84 for the highly B-doped films. Despite the high oxygen content, the highly B-doped films retained $Cu_4O_3$ along with CuO, indicating that boron modifies the oxidation pathway and stabilizes mixed-valence phases. This observation is supported by XPS analysis, which revealed increased concentrations of B–O species, defect-related oxygen, and mixed $Cu^{+}/Cu^{2+}$ states. The combined EDS, XPS, and XRD results suggest that boron promotes oxygen incorporation while simultaneously increasing the fraction of oxygen associated with defect-rich B–O environments, thereby suppressing complete conversion to CuO.

### 3.4. Film morphology

**Figure 9** illustrates the SEM images of Cu–O thin films, both undoped and B-doped deposited under varying oxygen partial pressures. The films exhibit a continuous, uniform surface without visible cracks, indicating high-quality films under the chosen deposition conditions. In the undoped films (**Figure 9a–c**), the surface is composed of densely packed grains with a fairly uniform size distribution. As oxygen partial pressure increases, grain size decreases slightly, resulting in a more compact, smoother morphology. Incorporating boron (**Figure 9d–i**) leads to a distinct alteration in surface morphology. At lower boron concentration, the films exhibit finer grains and a more uniform texture than the undoped films. With higher boron concentrations, the surface becomes notably smoother and denser, with fewer visible grains and some featureless areas. This suggests the development of a more disordered or amorphous-like surface layer, likely due to boron–oxygen bonding.

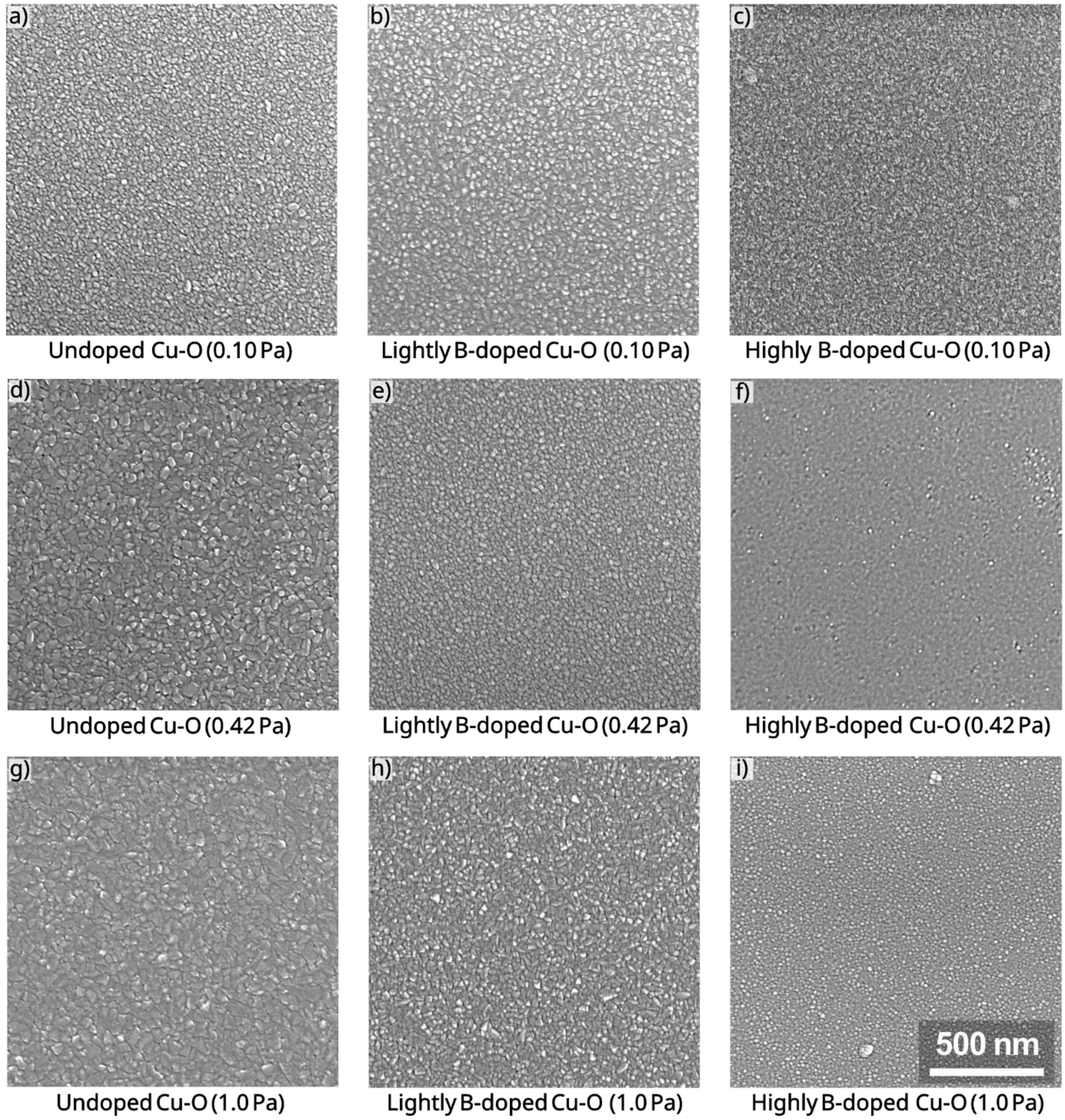

Figure 9 SEM micrograph of undoped Cu─O (a, d, g), lightly B-doped Cu─O (b, e, h), and highly B-doped Cu─O (c, f, and i) films at various oxygen partial pressures. The scale for all the images is same as shown in bottom right image.

### 3.5. Electrical and optical properties

**Figure 10a** illustrates the variation of the direct optical bandgap ($E_g$) for undoped, lightly B-doped, and highly B-doped Cu–O thin films as a function of oxygen partial pressure $p_{ox}$. At the lowest oxygen pressure of 0.1 Pa, a distinct widening of the bandgap occurs with increasing boron concentration, where $E_g$ expands from 2.47 eV for the undoped Cu-O film to 2.55 eV and 2.61 eV for the lightly and highly B-doped films, respectively. The phase evolution of the films directly justifies this widening. Higher boron concentration stabilizes the wider-bandgap $Cu_2O$-rich phase over the narrower-bandgap $Cu_4O_3$ phase [14,48,49]. As $p_{ox}$ increases from 0.1 Pa to 0.23 Pa, all films exhibit a sharp decrease in their optical bandgap values. This sudden reduction marks the structural transition from $Cu_2O$ towards the narrower-bandgap CuO phase. Beyond $p_{ox}$ of 0.35 Pa, the bandgap values stabilize and exhibit a gradual, minor decline. In this high-oxygen partial pressure regime (0.42 to 1.0 Pa), the values converge closely around 2.19 eV to 2.23 eV, with the high-B-doped film consistently maintaining the lowest bandgap (~2.19 eV) at $p_{ox} \geq 0.7$ Pa, as the low-bandgap CuO phase becomes dominant across all compositions [8,12,48].

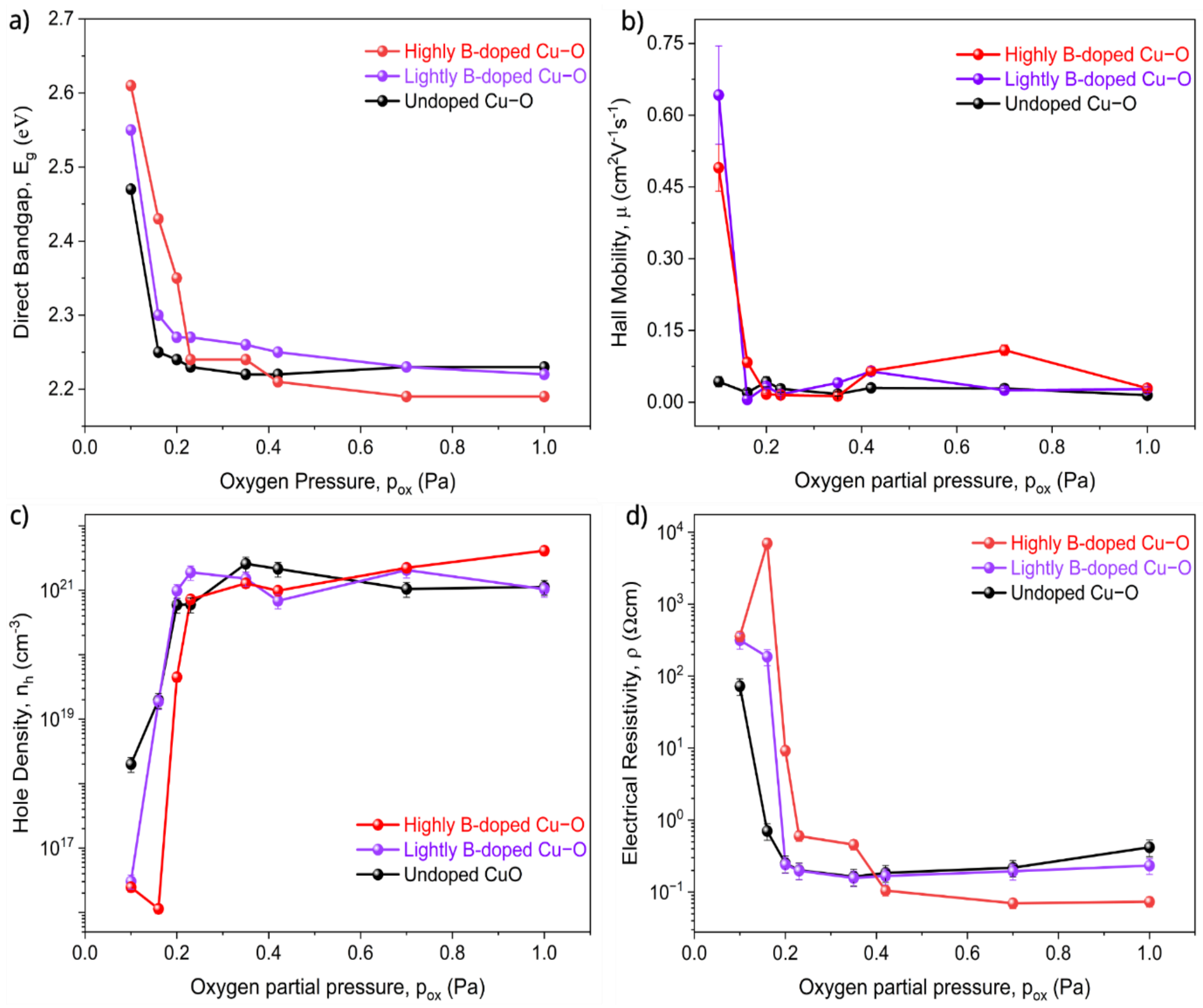

Figure 10 Variation of the (a) direct optical band gap ($E_g$), (b) Hall mobility (μ), (c) hole carrier concentration ($n_h$), and (d) electrical resistivity (ρ) of undoped and B-doped Cu-O thin films as a function of oxygen partial pressure.

The electrical transport properties of the undoped Cu-O thin films across varying oxygen partial pressures are displayed in **Figure 10b–d**. At a low $p_{ox}$ of 0.10 Pa, the undoped film exhibit relatively low carrier concentration ($n_h$) of approximately $2\times10^{18}$ $cm^{-3}$, a low Hall mobility ($\mu$) of 0.043 $cm^2V^{-1}s^{-1}$, and a corresponding electrical resistivity ($\rho$) of about 73 Ωcm. This high resistivity is characteristic of the mixed $Cu_4O_3$ (primary) + $Cu_2O$ phase present at this pressure. When the oxygen pressure is raised slightly to 0.16 Pa, the resistivity drops sharply by two orders of magnitude to ~ 0.71 Ωcm. This drop is driven by a massive, sharp increase in hole density, which climbs rapidly to exceed $5\times10^{20}$ $cm^{-3}$ at 0.23 Pa and peaks at $2.6\times10^{21}$ $cm^{-3}$ near 0.35 Pa. These dramatic changes in electrical properties correlate with the phase transition to single-phase CuO, which promotes a high density of native copper vacancy acceptors [24,49,50] at higher $p_{ox}$, where the films are predominantly CuO according to the XRD results, the carrier concentration remains in the range of $10^{21}$ cm$^{-3}$, while the resistivity gradually increases to approximately 0.42 Ωcm at 1.0 Pa. This increase in resistivity can be attributed to enhanced carrier scattering and reduced transport efficiency in the fully oxidized CuO phase [50–53].

The lightly B-doped Cu–O films exhibit electrical behavior similar to that of undoped films but with improved transport properties. At $p_{ox}$ = 0.10 Pa, the mobility reaches the highest value among all samples (~0.64 cm$^2$V$^{-1}$s$^{-1}$), while the carrier concentration remains relatively low (~$10^{17}$ cm$^{-3}$), resulting in a resistivity of approximately 300 Ωcm. As the oxygen partial pressure increases, the carrier concentration rapidly increases to the order of $10^{21}$ cm$^{-3}$, while the mobility decreases significantly. Consequently, the resistivity drops sharply and stabilizes around 0.15–0.25 Ωcm over a wide range of oxygen pressure. The relatively stable resistivity at higher oxygen partial pressures suggests that low boron incorporation effectively enhances carrier generation while maintaining moderate transport properties. This enhancement in electrical conductivity can be correlated with the phase evolution observed in XRD. The XPS results revealed the presence of B–O-related defect states and mixed Cu valence states that contribute additional acceptor-like defects and facilitate p-type conductivity [50,54]. Although CuO becomes the dominant phase at higher oxygen pressures, the electrical properties remain superior to those of undoped films, indicating that boron-induced defect states have a significant contribution in charge transport.

The initial increase in resistivity observed for the highly B-doped films between ($p_{ox}$=0.10) and 0.16 Pa is likely related to the boron-induced broadening of the $Cu_2O$/$Cu_4O_3$ phase stability window. Unlike in undoped and lightly doped films, where the films oxidized more rapidly with increasing oxygen partial pressure, the higher boron concentration stabilizes $Cu_2O$ and delays the formation of more-oxidized mixed-valence phases. Consequently, the film remains in a transitional regime at $p_{ox}$=0.16 Pa, where the introduction of $Cu_4O_3$ and associated structural disorder increases carrier scattering, while the carrier concentration has not yet increased to compensate for this effect. Upon further increasing $p_{ox}$ to 0.20 Pa, $Cu_4O_3$ becomes the dominant phase and a larger concentration of mixed $Cu^+/Cu^{2+}$ states and oxygen-related defects is established, resulting in enhanced carrier generation and a subsequent reduction in resistivity. Therefore, the delayed decrease in resistivity in the highly B-doped films closely follows the delayed phase evolution observed by XRD and reflects the strong influence of boron on the oxidation kinetics and defect chemistry of the Cu–O system. The mobility remains low but relatively stable after the initial decrease, indicating that the enhanced conductivity is primarily governed by the large increase in carrier concentration. This behavior correlates with the unique phase evolution of the highly B-doped films. XRD and Raman patterns show that these films evolve through $Cu_2O \rightarrow Cu_2O + Cu_4O_3 \rightarrow CuO + Cu_4O_3$, indicating that the mixed valence $Cu_4O_3$ phase remains stable over a broad range of $p_{ox}$. Since $Cu_4O_3$ contains $Cu^+$ and $Cu^{2+}$ ions, its stabilization could promote hopping conduction through mixed valence charge transfer pathways [14,50]. Furthermore, XPS analysis confirmed increased defect oxygen and the presence of B–O and possible B-O-Cu bonding configurations, which provide additional defect-assisted conduction channels [14,15,28,54–56].

The comparative analysis of all films clearly demonstrates that boron incorporation significantly modifies the optical and electrical transport properties of Cu–O films. In the low oxygen pressure regime ($p_{ox} < 0.16$ Pa), increasing the boron concentration broadens the $Cu_2O$-containing regions. While all films exhibit a reduction in band gap with increasing oxygen partial pressure due to the progressive formation of CuO, boron doping shifts the phase evolution toward higher $p_{ox}$ and broadens the stability range of mixed-valence copper oxide phases [14,48,49]. The lightly B-doped films exhibit improved electrical conductivity compared with the undoped films over a broad $p_{ox}$ range, whereas the highly B-doped films show the most pronounced enhancement in carrier concentration and the lowest resistivity. Combining the XRD, EDS, XPS, and Hall-effect results suggests that boron is incorporated primarily as B–O species [42], which modify the oxygen chemistry, stabilize defect-rich mixed-valence Cu–O phases, and promote charge transport [22,52,53,56–58]. As a result, the highly B-doped films achieve the best overall electrical performance, with a minimum resistivity of approximately 0.06 Ωcm, while simultaneously maintaining a tunable optical band gap in the range of 2.19–2.61 eV.

DFT calculations further support the experimentally observed reduction in resistivity in highly B-doped Cu–O films. The calculated density of states (DOS) shows that boron incorporation, particularly in the presence of nearby Cu vacancies, introduces additional electronic states near the Fermi level in both CuO and $Cu_2O$ lattices. This effect is more pronounced in CuO, where vacancy-assisted boron incorporation significantly narrows the band gap and increases the DOS near the Fermi level, thereby enhancing carrier availability and improving electronic transport.

## 4. Conclusions

The structural, chemical, and electrical properties of sputter-deposited Cu–O-B thin films were strongly influenced by oxygen partial pressure and boron incorporation. XRD and Raman analyses revealed a systematic phase evolution from $Cu_2O$/$Cu_4O_3$ to CuO with increasing $p_{ox}$, whereas boron doping broadened the stability window of $Cu_2O$ and $Cu_4O_3$ phases and delayed the complete formation of CuO toward higher oxygen partial pressures.

XPS analysis confirmed that boron exists predominantly in the oxidized form as B–O species, leading to increased defect-related oxygen and to stabilization of mixed $Cu^{+}$/$Cu^{2+}$ valence states. The formation of B–O and possible B–O–Cu bonding configurations altered the local oxygen chemistry and promoted mixed-valence copper oxide phases.

The electrical properties were also strongly affected by boron incorporation. While undoped films showed an increase in resistivity from 0.13 Ωcm (at $p_{ox}$=0.35Pa) to 0.3 Ωcm (at $p_{ox}$=1.0 Pa) due to the dominant formation of CuO, highly B-doped Cu–O films exhibited a reduction in resistivity across the investigated oxygen pressure range. The lowest resistivity achieved was approximately 0.06 Ωcm, which is probably attributed to defect-assisted transport and mixed-valence conduction pathways induced by boron incorporation.

Overall, boron doping proved to be an effective approach for tuning the phase composition, defect chemistry, and electrical conductivity of Cu–O thin films, making these materials promising for optoelectronic and photovoltaic applications.

**Declaration of Generative AI and AI-assisted technologies in the writing process**

During the preparation of this work the authors used Grammarly, ChatGPT, and Paperpal to improve the grammar and style of the draft text. After using this service, the authors reviewed and edited the content as needed and take full responsibility for the content of the publication.

**Declaration of competing interest**

The authors declare that they have no known competing financial interests or personal relationships that could have appeared to influence the work reported in this paper.

**Acknowledgments**

This work was supported by the project Quantum materials for applications in sustainable technologies (QM4ST), funded as project No. CZ.02.01.01/00/22_008/0004572 by Programme Johannes Amos Comenius, call Excellent Research.